\documentclass[preprint,11pt,5p,twocolumn]{elsarticle}
\journal{Ultramicroscopy}

\usepackage{amssymb}
\usepackage{todonotes}
\usepackage{fixltx2e}
\usepackage{txfonts}
\usepackage{placeins} %for %\FloatBarrier
\usepackage{flushend}
\begin{document}

\begin{frontmatter}

\title{Sub-100 nanosecond temporally resolved imaging with the Medipix3 direct electron detector}
%\tnotetext[label0]{This is only an example}

\author[label1]{Gary W. Paterson\corref{cor1}}
\address[label1]{SUPA, School of Physics and Astronomy, University of Glasgow, Glasgow G12 8QQ, United Kingdom}
\ead{Gary.Paterson@glasgow.ac.uk}

\author[label1]{Raymond J. Lamb\corref{cor1}}
\ead{Raymondlamb91@gmail.com}

\author[label2]{Rafael Ballabriga}
\ead{rafael.ballabriga@cern.ch}
\address[label2]{CERN, 1211 Geneva 23, Geneva, Switzerland}

\author[label1]{Dima Maneuski}
\ead{Dima.Maneuski@glasgow.ac.uk}

\author[label1]{Val O'Shea}
\ead{Val.OShea@glasgow.ac.uk}

\author[label1]{Damien McGrouther\corref{cor1}}
\ead{Damien.McGrouther@glasgow.ac.uk}

\cortext[cor1]{Corresponding author}

\begin{abstract}
Detector developments are currently enabling new capabilities in the field of transmission electron microscopy (TEM).
We have investigated the limits of a hybrid pixel detector, Medipix3, to record dynamic, time varying, electron signals. 
Operating with an energy of 60~keV, we have utilised electrostatic deflection to oscillate electron beam position on the detector. 
Adopting a pump-probe imaging strategy, we have demonstrated that temporal resolutions three orders of magnitude smaller than are available for typically used TEM imaging detectors are possible. 
Our experiments have shown that energy deposition of the primary electrons in the hybrid pixel detector limits the overall temporal resolution. 
Through adjustment of user specifiable thresholds or the use of charge summing mode, we have obtained images composed from summing 10,000s frames containing single electron events to achieve temporal resolution less than 100~ns. 
We propose that this capability can be directly applied to studying repeatable material dynamic processes but also to implement low-dose imaging schemes in scanning transmission electron microscopy.

\end{abstract}


\end{frontmatter}

\section{Introduction}
\label{sec1}
Enabled by high coherence electron sources, advances in aberration corrected electron optics~\cite{Haider_nature_1998, pennycook_2006} and high stability power supplies, modern transmission electron microscopes (TEM) provide images with spatial resolution exceeding the interatomic spacing in materials.
These capabilities have provided tremendous insight into fundamental materials physics and structure-property relationships. 
However, the functional performance of advanced materials also depends on their response to time changing conditions. 
In this respect, conventional TEMs with continuous current electron sources are much more limited in providing insightful dynamic information. Their time resolution can primarily be limited by the image detection technology employed but more fundamentally by the brightness of the electron source. Typically available electron beam currents, ${10\mathrm{s\, nA}}$ at most, ultimately limit the signal-to-noise ratio (SNR) in images with frame times of ${\mathrm{10s-100s\,  \muup s}}$ assuming that acquisition is possible over such short durations and at high rates~\cite{Beacham_JOI_2011}.

A very effective route to achieving time-resolved imaging has been the implementation of photo-emission electron sources driven by lasers.
This has enabled the specialised field of ultra-fast electron microscopy (UEM) in which the illumination duration is of the order of the femtosecond laser pulse duration~\cite{PIAZZA_chem_phys_2013, Kieft_struc_dyn2_2015, BUCKER20168, Kuwahara_APL_2016, FEIST_2017}.
Utilising a pump-probe imaging methodology for the study of repeatable phenomena has led to insights in areas such as nanophotonics~\cite{Feist_nat_2015_nearfield}, atomic structural dynamics~\cite{Gaolong_sci_rep_2015}, magnetic dynamics~\cite{da_Sliva_2018}, and even electron dynamics~\cite{Hassan_nature_2017_electron_dynamics}.
Stochastic, non-repeatable, processes have also been studied using high intensity, nanosecond duration laser pulses in dynamic TEM (DTEM).
In single-shot mode imaging, large numbers of electrons are photo-emitted from the source, traversing the column as a single bunch~\cite{BOSTANJOGLO2000141, LAGRANGE_2008_dtem, BROWNING201223}.
Across both UEM and DTEM techniques, the wide use of thermionic emitters leads to spatial and temporal coherence of the electron bunches being significantly lower than routinely obtained from conventional continuous electron sources~\cite{Kuwahara_APL_2016, Ji_cathode_2017}.
This has limited time-resolved imaging to nanometre spatial resolutions.
Photoemission from field emission gun sources is being investigated and holds potential for improvements~\cite{FEIST_2017, Houdellier_Ultramicr_2018}. 

In this article, we report the feasibility of utilising a direct, pixelated counting detector, Medipix3~\cite{CSM,Ballabriga_JInstr_2011} on an unmodified, continuous current source TEM to implement `pump-probe' imaging.
Having recently demonstrated that the single electron sensitivity of these detectors provides significant advantages for lower-energy (60-80~keV) TEM imaging~\cite{MIR_UM_2017_medipix} and 4-D scanning TEM (STEM) imaging~\cite{PIXDPC}, we demonstrate that around three orders of magnitude improvement in temporal resolution can be obtained when operating using continuous beams.

The image detector is based upon a hybrid pixel architecture where a semiconductor sensor (typically silicon, 300-500~${\muup}$m thick) is directly bump bonded to the Medipix3 application specific integrated circuit (ASIC).
The ASIC is pixelated, containing both analogue and digital circuitry repeated upon a 55~$\muup$m pitch.
Applied in electron microscopy, we have demonstrated that the user adjustable event energy threshold allows counting of incident primary beam electrons without influence from thermal detector noise.
Readout of pixels is also noiseless (through in-pixel digital shift registers) which is a critical feature for `pump-probe' imaging based on the summation of many thousands of frames.
According to the beam currents typically available in TEMs with continuous sources, frame exposure times of 10 microseconds or less will be composed of multiple single electron events.

The physical deposition of primary electron energy, the sensor charge currents that result, the chosen energy threshold and subsequent processing by the pixel circuitry all dictate the detector's resultant modulation transfer function (MTF), detector quantum efficiency (DQE) and time resolution performance. 
Following an in-depth discussion of relevant aspects of the pixel circuitry, the following sections report characterisation of our time varying signal in the TEM before going on determine the time resolution of the Medipix3 detector. 
We find that the pixel energy threshold may be used to improve the temporal resolution available in the single pixel mode of operation. 
Related to this, we have also investigated how the use of the charge summing mode of operation~\cite{CSM} also improves temporal resolution and affects the delay between an electron hit and the subsequent digital count being registered. 
Finally, we present in-situ measurements of a dynamic electron beam process, demonstrating stroboscopic imaging in the sub-100 nanosecond regime in an unmodified TEM with a SNR well beyond the Rose criterion~\cite{Rose}.

\section{Medipix3 hybrid pixel architecture and temporal imaging}
\label{sec2}
A significant factor in the temporal response of the Medipix3 detector is how the secondary electron-hole pair charge generated in the sensor layer is processed by the analogue sections of the pixel circuitry.
We, therefore, begin by briefly summarising the pixel architecture before going on to discuss the operation of the analogue components and how their response varies with the range of interactions of the primary electrons.

\begin{figure*}[!hbt]
    \centering
        \includegraphics[width=16cm]{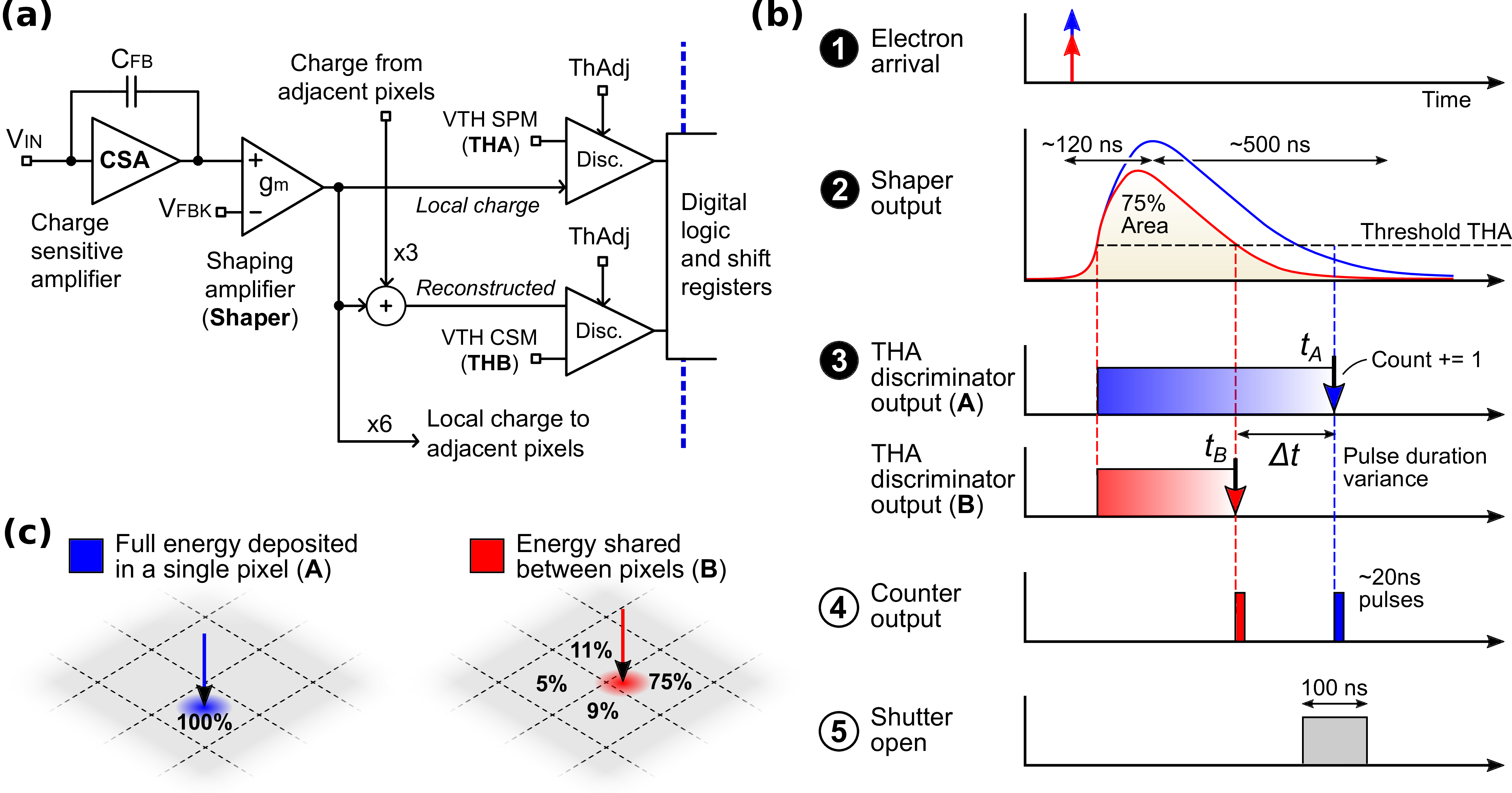}
        \caption{(a) Medipix3 analogue front end. (b) Simplified schematic of the (1-3) analogue and (4-5) digital stages of electron detection in the Medipix3 architecture for the two cases shown in (c). The variance in discriminated pulse time above threshold THA, $\Delta t$, is a result of electrons depositing their energy in one pixel (case A) and across multiple pixels (case B), and is the primary limiting factor in the time resolution of the Medipix3 detector.}
    \label{fig:PixelResponce}
\end{figure*}

The detector chip that we investigated featured a 300~$\muup$m thick silicon sensor bump bonded to a single Medipix3 ASIC with a 256$\times$256 pixel array.
The pixel pitch was 55~$\muup$m. 
The silicon surface is coated with an aluminium thin film which acts as a top electrode when applying a biasing voltage across the sensor (+90~V was used here).
The applied bias generates an electric field inside the sensor that causes the positive charge carriers (holes) to move towards the pixel electrodes that interface it to the ASIC, and electrons to move in the opposite direction.
The movement is by drift under the electric field and also by diffusion. 
This moving charge in the sensor volume induces a signal in the analogue section of the pixel electronics, as depicted in Figure~\ref{fig:PixelResponce}(a), and is processed according to the mode of operation.
Operating in single pixel mode (SPM), a voltage signal proportional to the charge collected by each pixel (which is proportional to the amount of energy deposited by the primary electron) is compared to user specified threshold voltages.
If only a lower energy threshold is utilised, then when this is exceeded, a pulse is sent to the digital pixel circuitry instructing that an event be counted.
In the charge summing mode (CSM)~\cite{CSM} of operation, links enable the summing (reconstruction) of charge spread over neighbouring pixels in 2$\times$2 pixel clusters.
The reconstructed energy is assigned to the pixel which obtained the largest energy deposition. 
The decision of as to which pixel the signal should be assigned is based on the local signal deposited in the pixels.
The reconstruction operation is performed in the analogue section and the primary electron `hit' pixel assignation is done by digital arbitration logic based on the timing of the signals.
CSM was originally conceived to handle multi-pixel events arising from spreading of the secondary electron-hole charge under diffusion. 
For high energy electrons, we have shown that the algorithm can provide simultaneous improvement of both the DQE and MTF for energies in the range 60-80~keV~\cite{MIR_UM_2017_medipix}, therefore correcting for lateral spreading in the sensor of the primary radiation also.

The process of detecting and counting an incident primary electron is now discussed for two cases with close reference to Figure~\ref{fig:PixelResponce}.
Case A (depicted in blue) in Figures~\ref{fig:PixelResponce}(b) and (c) represents an idealised electron event where 100~$\%$ of the primary electron's energy is deposited within the collection area of a single pixel. 
From the point of view of timescales in the overall counting process, the initial impingement and energy deposition of the primary electron in the silicon sensor may be viewed as a prompt, instantaneous starting event. 

Following the electron arrival, e-h pair charge generated in the sensor volume along the path travelled by the primary electron moves under the applied bias and this movement induces a signal at the readout pixel electrodes.
The charge signal is converted to a voltage signal by the charge sensitive amplifier (CSA) whose output is then fed to a shaping amplifier. 
The output of the Shaper is a current pulse with amplitude proportional to the collected charge, as illustrated in step 2 of Figure~\ref{fig:PixelResponce}(b). 
The rise time of the pulse is of the order of 120~ns, while the decay time is of the order of 500~ns~\cite{Llopart_IEEE_medipix2_2002}.
Two discriminators compare the amplitude of the shaped pulse with the user specified threshold levels. 
In the case where only the low threshold, THA, is specified then, step 3 of Figure~\ref{fig:PixelResponce}(b) shows the resultant output of the discriminator comparing the pulse to THA.
The discriminated pulse, of duration $t_A$, is processed by the digital pixel circuitry, which, provided the image shutter signal is high (open), then emits a short ($\sim$20~ns) pulse on the falling edge of the discriminated pulse, incrementing the active counter.

The blue curves of case A in Figure~\ref{fig:PixelResponce}(b) apply when primary electrons deposit their entire kinetic energy within a single pixel.
This only tends to occur when electrons impinge towards the centre of the pixel coordinate and their maximum lateral scattering is less than the distance to the nearest neighbouring pixel.
Monte Carlo~\cite{casino_2007} simulations show that 95$\%$ of incident 60~keV electrons deposit all energy within a radius of 8~$\muup$m.
Therefore, by geometric calculation, a 55~$\muup$m pixel has an area $\sim$50\% of its total in which electron impingement will deposit all their energy into that single pixel.
Conversely, for impingement away from the pixel centre, due to lateral scattering, it is likely that secondary e-h pairs will be registered not only by the pixel of incidence but also in adjacent pixels, sharing out the primary energy.
The scenario of an incident electron sharing energy across multiple pixels is shown by the red data in case B of Figures~\ref{fig:PixelResponce}(b) and (c).

\begin{figure*}[htb!]
    \centering
        \includegraphics[width=17.5cm]{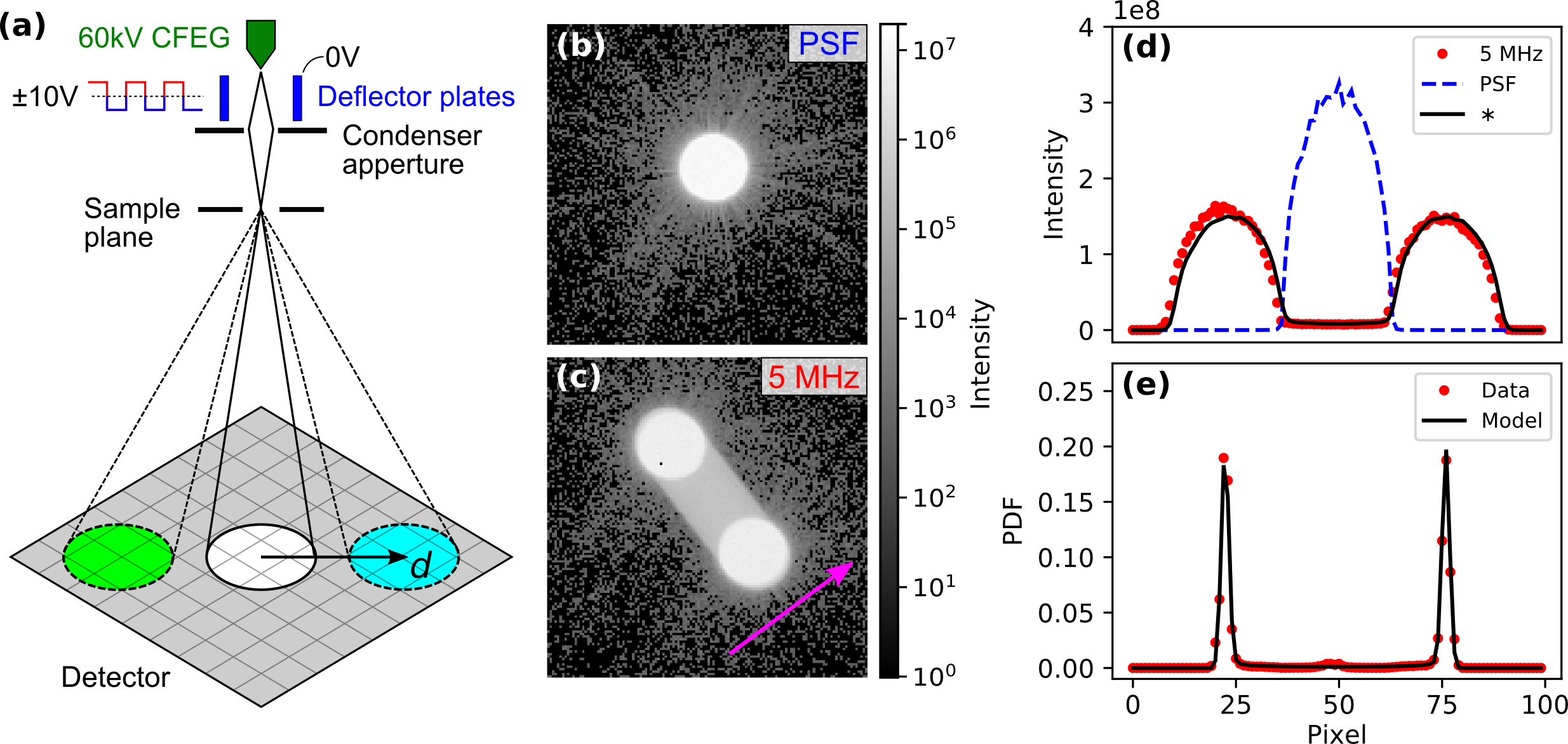}
        \caption{Schematic of the deflection plates and detector arrangement (a) showing the static beam (white) and effect of a square wave applied to deflector plates (green, cyan). Long exposures of the beam while (b) stationary and (c) during application of a 5~MHz square wave. (d) Summed profiles of the long exposure images, representing a point spread function (PSF) from (b) and an intensity distribution from (d). The sum direction is shown by the magenta arrow in (c). The probability distribution function (PDF) obtained through deconvolution is shown as symbols in (e). The result of fitting a model to the PDF is shown as a black line in (e). The convolution ($\ast$) of this with the PSF is shown as a black line in (d) and matches the data (red symbols) well.}
        \label{fig:Plates}
\end{figure*}

The analogue output pulse from the Shaper will possess a smaller peak voltage and smaller integrated area in case B, with the difference compared to case A equal to the charge shared to adjacent pixels (25\% in this example).
Consequently, the amount of time the Shaper pulse is above the THA voltage threshold will be shorter, hence creating a discriminated pulse of duration $t_B<t_A$.
Thus, variation in spatial location of impingement of primary electrons with reference to the pixel centres leads to a significant variation in energy and, hence, $\Delta t$ in the discriminated pulse length that ultimately leads to event counting. 

Important to whether an event is registered to a particular image frame or not is the shutter signal (step 5 of Figure~\ref{fig:PixelResponce}(b)).
The shutter signal timing depends on the clock rate (up to 200~MHz) supplied to the Medipix3 by it's attached readout system.
In our experiments we used a Merlin readout system (from Quantum Detectors Ltd) with a clock rate of 120~MHz~\cite{Horswell_2011_joi_100MHz}. 
According to the Merlin specifications, shutter times have a jitter of 20~ns~\cite{Plackett_2013_joi_jitter}. 
Thus, it would seem feasible to reliably specify minimum shutter (frame) times of 50-100~ns in duration.  
This minimum shutter timescale is significantly shorter than the lengths of the Shaper pulse and resultant discriminated pulse ($>$few hundred ns) and gives rise to two important consequences.
Firstly, when primary electrons are detected, they have arrived at the detector before the shutter signal has made it's closed-open (low-high) transition.
Effectively, the detector sees backwards in time by a period of hundreds of nanoseconds.
Secondly, it is clear that for two monoenergetic primary electrons, such as the cases A and B depicted in Figure~\ref{fig:PixelResponce}, arriving simultaneously at well separated pixels, due to the variation in discriminated pulse length, $\Delta t$, both events are unlikely to be detected within the same short shutter period (\emph{i.e.} not within the same image frame, as drawn in the figure).
Therefore, the total range of the variation in the discriminated pulse length is expected to dictate the limit on the minimum achievable time resolution of Medipix3.

\section{Experimental results}
\subsection{Dynamic illumination}
\label{sec3}
In order to determine the time response of the Medipix3 detector for a given mode of operation, a dynamic object of known and repeatable properties must be imaged.
The use of deflector plates to control sample dose is a popular method in biological fields where beam-sensitive specimens are commonplace~\cite{Hayashida2007505, Ohtsuki1975163}.
Here, we use a similar method to create a dynamic illumination source with which to control the position and dwell time of an electron beam on the detector.
In this approach, the deflection plates do not gate the exposure; instead, they provide a periodic dynamic image.
In effect, the beam is our `sample' and the perturbation of its position is the `pump' and the detector exposure is the `probe' of the pump-probe methodology.

The experiments were performed in a JEOL ARM200cF~\cite{McVitie_Ultramicr} equipped with deflector plates located directly after the electron gun and before the primary optics, as shown in the schematic of Figure~\ref{fig:Plates}(a).
The microscope was operated at 60~kV in STEM mode without a sample present, producing an image of the circular condenser aperture in reciprocal space, as shown in panel (b).
As only relatively small deflections in the beam are needed for our purpose, the $\pm$10~V output of a 50~$\Omega$ output impedance Agilent 33250A 80~MHz waveform generator was used to drive one deflection plate with a square wave while the other remained grounded.
The 10-90\% rise and fall time of the waveform generator itself was measured to be 5.6$\pm$0.5~ns at frequencies of 100~kHz and 5~MHz.
In the experiments, the capacitance of the cabling, of the deflector plates (14.5~pF at 100~kHz), and parasitic impedances will influence the response of the beam, so it is important to characterise it experimentally.

To determine the response time of the illumination source, long exposure images were taken with a static beam and with the deflector plates excited with a 5~MHz square wave, as shown in Figures~\ref{fig:Plates}(b) and \ref{fig:Plates}(c).
The perturbation effectively splits the undeflected beam into two through inducing an angular tilt in the specimen plane. This results in a spatial separation by a distance, $d$  in the diffraction plane which was projected onto the detector.
The intensity in the gap between the two disks is caused by the transition of the beam from one deflected position to the other.
The time of the deflection transition is dependent on the rise and fall time of the signal generator and the time constant, $\tau = RC$, of the deflector plates, cabling and source, where $R$ and $C$ are the equivalent resistance and capacitance of the system.
For our setup, $\tau$ should be around or below-1~ns, so we can combine its effect with the larger rise time of the source to estimate a single transition time.

The image of the static beam may be regarded as a point spread function (PSF) and used to calculate the relative time spent at each position during the beam transition.
Since our system is effectively one-dimensional (1-D), we reduce the influence of noise, and simplify the analysis, by summing the pixels along the direction perpendicular to the deflection axis after rebinning the data by a factor of two.
Figure~\ref{fig:Plates}(d) shows the 1-D profiles created by this procedure, with the 5~MHz stimulus long exposure data shown as a red symbols and the static beam shown as a blue dashed line. 
The probability distribution function (PDF) during excitation of the deflection plates is extracted by deconvolving the 5~MHz profile with the PSF.
The result of doing this using 256 iterations of a Richardson Lucy algorithm~\cite{Richardson:72, Lucy:1974yx} is shown as symbols in Figure~\ref{fig:Plates}(e).
Deviations of this profile from two delta functions represents the finite beam transition time and other sources of beam movements.

With the rise time as the dominant mechanism of broadening, we approximate the transition profile as an error function.
The width, $\sigma$, of the equivalent Gaussian that would be used to convolve a perfect square wave is 0.39$\times$ the 10-90~\% rise time.
The solid line in Figure~\ref{fig:Plates}(e) is a fit to the data of such a function.
Additional broadening was required to obtain a good fit and this was approximated by introducing a small Gaussian convolution into the modelled PDF.
The sources of such broadening are most likely to be ringing and reflections from impedance mismatches.
To confirm that the model is accurate, we convolved the PDF with the PSF \& Gaussian broadening and compare this to the 5~MHz experimental profile.
The reconstructed profile is shown as a solid black line in Figure~\ref{fig:Plates}(d) and agrees very well with the original data (red symbols).
The 10-90~\% rise time extracted following this procedure was 7$\pm$1~ns, equivalent to $\sigma = 2.7$~ns.
This value sets an upper limit to the effective lifetime of the 5~MHz dynamic source of $\sim$90~ns.

With the response of the deflection plates and driving circuitry determined, we will use this apparatus to test the Medipix3 detector time resolution limits using a stroboscopic technique.
We will show in the next section that the resolution achievable can be maximised by optimising the detector parameters to take account of the Medipix3 pixel response discussed in Section~\ref{sec2}.

\subsection{Optimising time resolution}
\label{sec4}
In Section~\ref{sec2} we identified the physical process connecting charge sharing and time resolution in the Medipix3 detector.
In this section we explore the influence of charge sharing on the variance in electron detection time through increasing the rejection of electrons that have experienced significant charge sharing in SPM by varying THA, and by employing charge reconstruction in the CSM mode of operation.

\begin{figure}[!ht]
    \centering
    \includegraphics[width=8.5cm]{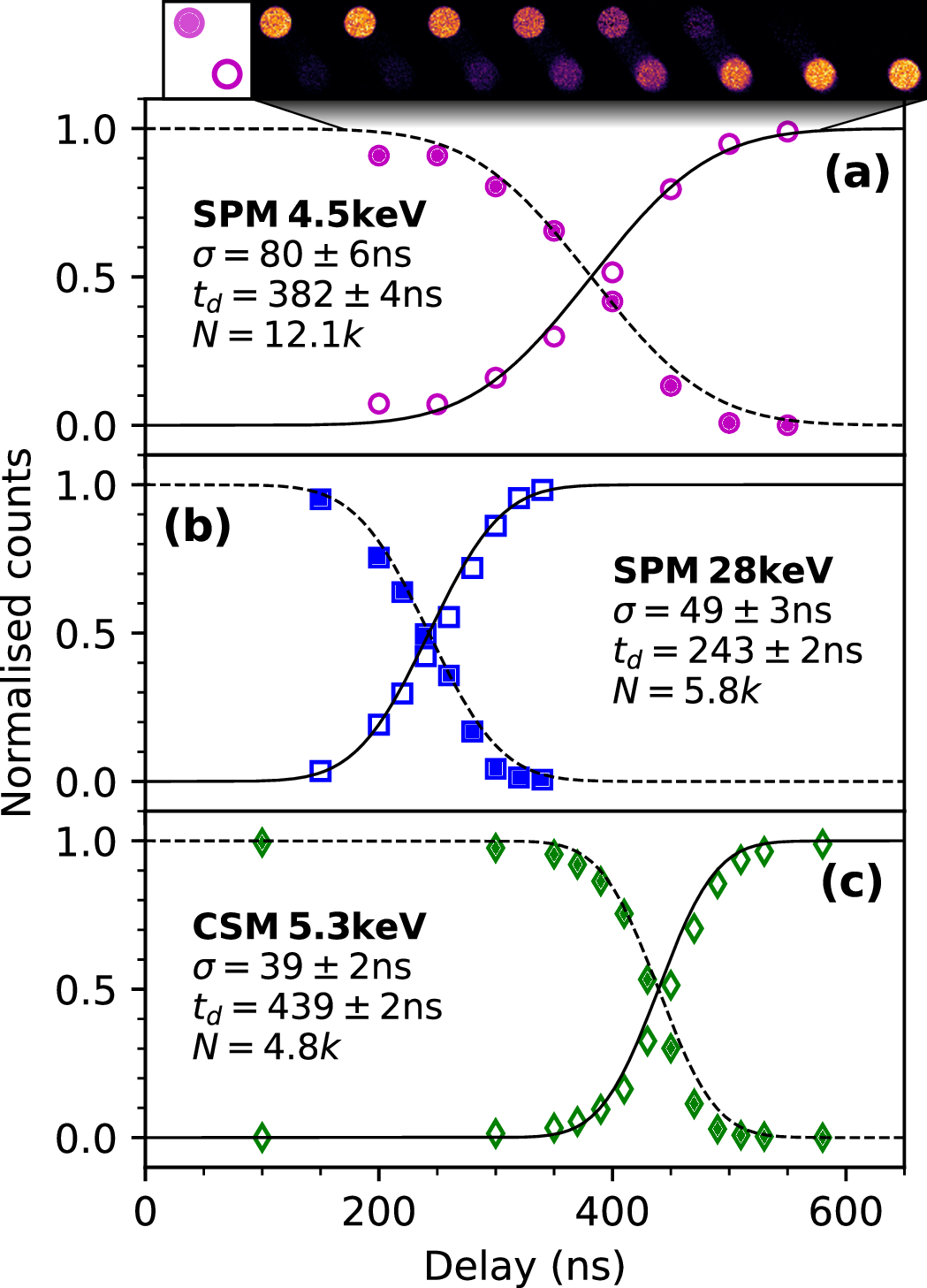}
        \caption{Normalised intensity of the deflected beams at different exposure delay times, showing the measured time response of an identical transition under different acquisition conditions. SPM acquisitions with a low and high threshold are shown in (a) and (b), while CSM data is shown in (c). The inset to (a) are images analysed in that experiment. The lines in (a-c) are fits to the data of error functions of width $\sigma$ and centre position $t_d$. $N$ is the average counts per data point.}
    \label{fig:hs_mode}
\end{figure}

To experimentally determine the variance in the detection time under different modes of operation and thresholds, we stroboscopically recorded the detector response to a Heaviside step generated using the dynamic illumination source discussed in Section~\ref{sec3}.
A $\pm$10~V square wave with a period of 10~$\muup$s (f = 100~kHz) from the signal generator was applied across the deflector plates and the TTL from a second signal generator was used to introduce a variable delay on the triggering of the Medipix3 camera.
An image was acquired with a shutter exposure of 20~ns as a function of delay time over 50~k periods, and the resulting images summed for further analysis.
In these experiments, additional cabling was used, creating a longer beam transition time than in the optimised setup discussed in Section~\ref{sec3}.
This will increase the apparent detector response time of all results in this section, but will not affect the trends we find.

The results of this experiment are shown in Figure~\ref{fig:hs_mode}.
SPM data with 4.5~keV and 28~keV thresholds are shown in (a) and (b), respectively, while CSM data with a 5.3~keV THA threshold is shown in panel (c).
A threshold of of 4-5~keV marks the point just above the thermal noise floor.  
For the CSM data, THB was set to be equal to THA so that all single electron impacts were counted.
The open and closed symbols show the counts from regions of the normalised images encompassing the area of the pre- and post-transition beam, respectively, at acquisition delay times spanning the detection of the beam deflection.
An example of the summed images is shown in the inset to (a) for the detector in SPM mode with a low threshold voltage.
As the beam transitions from one position to the other, one spot fades while the other increases in intensity, with the normalised counts in each spot location transitioning from 0 to 1 and \emph{vice versa}. 
To characterise the detector response, we fit error functions simultaneously to each pair of intensity profiles in Figure~\ref{fig:hs_mode} to extract the mean delay, $t_d$, and the characteristic spread in response time, $\sigma$.
The fit results are shown as black lines and annotations in each panel of the figure, along with the mean counts per image, $N$.
We first consider the case of SPM.

\begin{figure}[!ht]
    \centering
    \includegraphics[width=7.5cm]{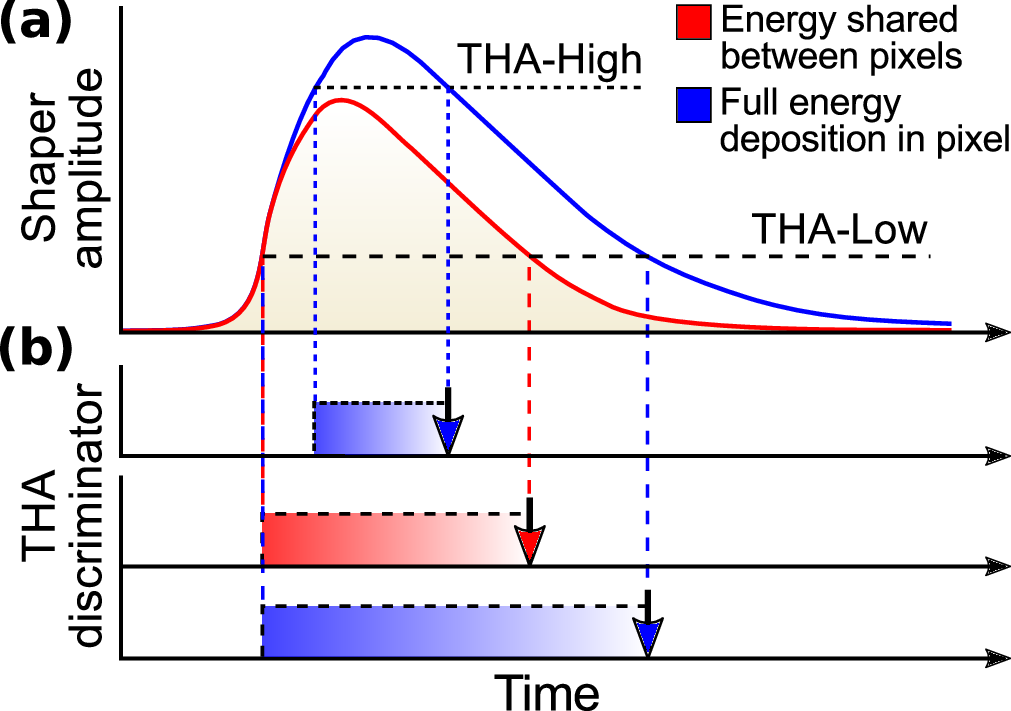}
        \caption{The effect on the delay \& variance in detection time of charge sharing on the (a) Shaper and (b) THA discriminator outputs for two different threshold voltages, THA-High and THA-Low. The peak voltage of the non-charge shared electron (blue curve) is higher than both thresholds and is accepted in both cases. The red curve, representing the charge shared electron, has a peak voltage below the higher threshold and is not registered as an electron count at the threshold value, reducing the delay and variance in detection time.}
    \label{fig:ThresholdHighLow}
\end{figure}

Increasing the threshold from 4.5~keV to 28~keV in SPM results in a reduction in $\sigma$ from 80~ns to 50~ns, indicating a smaller variance in electron detection duration; a 140~ns shorter average delay between the initial electron impact and the digital electron count; and a reduction in the total counts per integrated frame from an average of 12~k to 6~k.
These are all as a result of increasing the rejection of charge-shared electron impacts as depicted in Figure~\ref{fig:ThresholdHighLow}.

Figure~\ref{fig:ThresholdHighLow}(a) depicts the Shaper pulse response to electrons depositing all energy in a single pixel (shown in blue) and charge-shared electrons (shown in red) arriving at the detector at the same time.
As discussed previously, a characteristic of charge shared electrons is a reduction in the peak voltage output of the Shaper pulse.
Setting the threshold voltage above that generated by a charge shared electron will cause the detector to reject any electrons depositing an energy equal to or below this equivalent level.

The discriminator response is shown in Figure~\ref{fig:ThresholdHighLow}(b) for two levels of threshold, THA-High and THA-Low, for both electron energy sharing cases.
With a low threshold value (dashed lines in (a) and (b)), both electrons cause a Shaper pulse response above threshold and will be counted.
The time at which the count is registered occurs at the falling edge of the discriminator pulse, marked by vertical arrows and so, similar to the situation depicted in Figure~\ref{fig:PixelResponce}, there is a variance in the arrival of the two electrons being registered.

In the case of the high threshold level, only the electron depositing all of its energy within a single pixel will exceed the threshold and be counted. 
Events where there is charge sharing amongst pixels will not be counted.
From Figure~\ref{fig:ThresholdHighLow} it is clear that, for high threshold values, the duration of the discriminator pulse will be shorter when compared to low thresholds.
This has two effects.
First, shortened discriminator pulse lengths result in reduced pulse duration variance and hence better time resolution when counting over many events.
Secondly, the delay time, $t_{d}$, the time taken by the pixel to process the event leading to incrementation of the counter, is also shortened.

\begin{figure}[t]
    \centering
    \includegraphics[width=8.0cm]{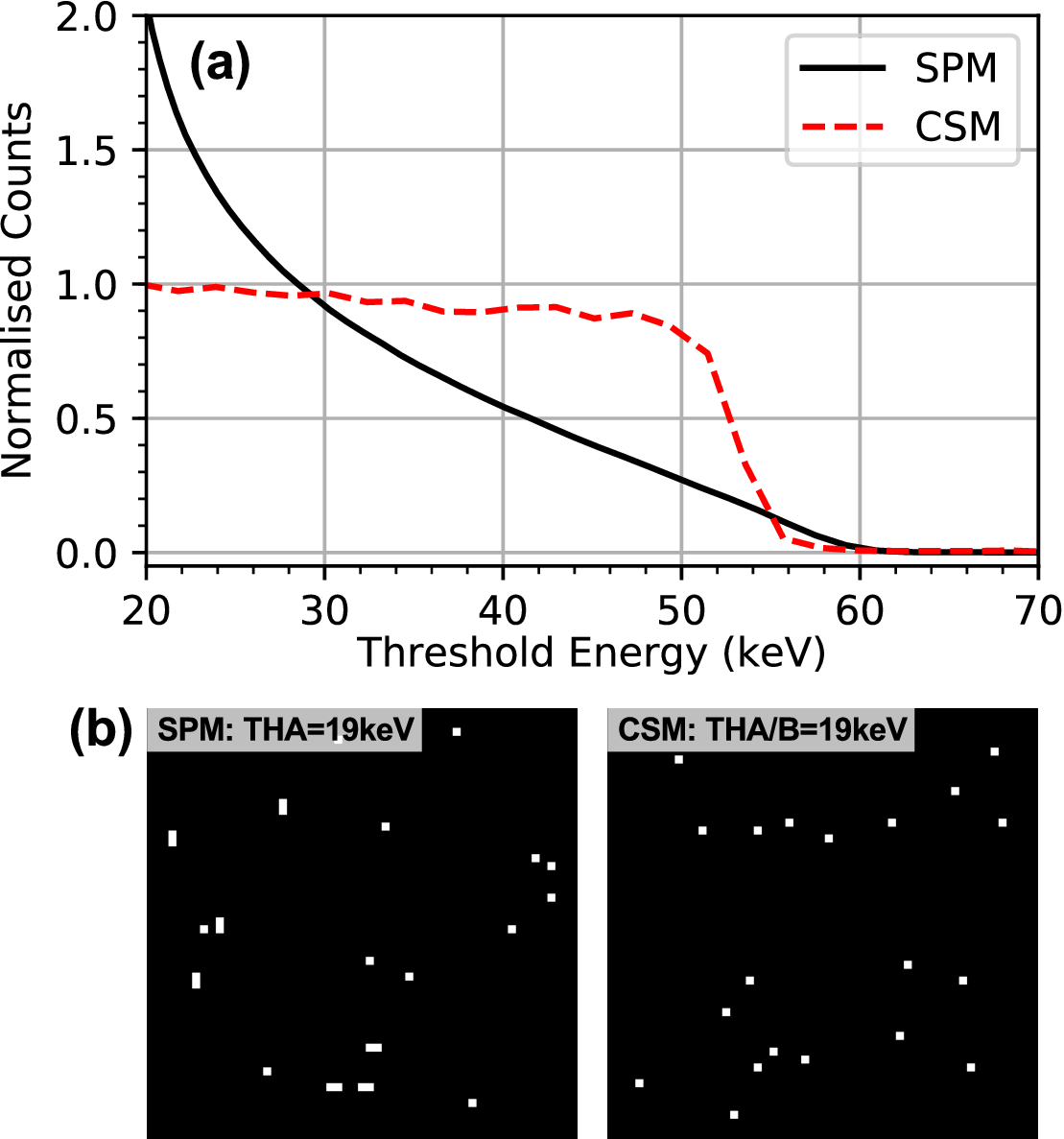}
        \caption{(a) The effect of varying threshold on the normalised counts measured in SPM and CSM with 60~keV electrons. In SPM, the threshold is THA. In CSM, the threshold is THB; THA was set to 5~keV. (b) Examples sections of individual frames produced from single 60~keV electron hits in SPM and CSM, showing the reduction in the energy spread when acquiring in CSM.}
    \label{fig:Threshold_Counts_Mode}
\end{figure}

In SPM with a low threshold, the mean electron cluster size for a single electron impact is much larger than 1, resulting in an inflated total number of counts.
At the higher threshold value used here, the mean number electrons per cluster is closer to 1.
Even higher threshold values may potentially be used to further improve the time resolution.
However, a significant disadvantage to doing this is that the overall fraction of impingement electrons that are actually counted and, hence the detector DQE, would be strongly reduced, as depicted in the black curve of Figure~\ref{fig:Threshold_Counts_Mode}(a) (discussed later).

Returning to Figure~\ref{fig:hs_mode}, the CSM curve in panel (c) shows that the response is delayed significantly to 440~ns but has a slightly faster transition ($\sigma = 40$~ns) than does the high threshold SPM data. 
In CSM, the charge deposited in pixels (provided it is greater than the THA threshold) is processed in ${2\times2}$ pixel blocks and the total charge assigned to the pixel with the greatest individual charge in a `winner takes all' design~\cite{CSM}. 
If the total charge in the `winning' pixel exceeds the second threshold, THB, then it is counted.
In this manner, electron events where charge sharing has occurred will be assigned to one pixel.
The increment of the winning pixel's count is delayed because it can only happen after the Shaper amplitude of all pixels in the relevant blocks drop below the low THA threshold and this is determined by the pixel with the most energy deposited in it. 
As a result, apparent completion of the beam transition in the CSM data occurs at a similar delay time to the low THA SPM data (\emph{c.f.} Figure~\ref{fig:hs_mode}(a) and (c)).
It is due to the CSM architecture that the Medipix3 is designed to trigger on the falling edge of the discriminator output.
Triggering on the rising edge has potential to improve the time resolution but is not possible with Medipix3.

When operating in CSM, it is sensible to set the THA threshold to a low value (so that all individual pixels in an event cluster are considered) and THB to a higher value.
Since the latter operates on summed charge, the DQE performance is improved, provided primary electron energies have an average lateral spread that is less than the ${2\times2}$ pixel block size~\cite{MIR_UM_2017_medipix}.
The count dependence on threshold value for CSM and SPM are shown in Figure~\ref{fig:Threshold_Counts_Mode}(a).
As THB in CSM is lowered through the beam energy, the normalised counts rise sharply from 0, reaching $\sim$0.7 by 50~keV, then continue to slowing increase towards $\sim$1.0 at 20~keV.
The weak dependence on threshold in CSM is a direct result of the charge reconstruction.
In contrast, the SPM counts smoothly and continuously increase with decreasing THA, passing a value of 1 at around half the beam energy, and reaching a value of $\sim$2 at 20~keV.
This artificially inflation of the number of counts at low thresholds is due to charge sharing between pixels giving rise to multiple counts per primary electron.
A typical example of the reduction cluster size produced from single primary electron impacts by enabling CSM is shown in Figure~\ref{fig:Threshold_Counts_Mode}(b).

As was discussed for SPM mode, reducing the variance in the duration of the discriminator pulse that leads to event counting is expected to improve the temporal resolution.
In CSM, the relevant discriminator pulse is based on comparison of summed charge to THB.
Thus, where primary electron energy and lateral spread permit proper CSM operation, it has the effect of reducing the variance in the Shaper pulse width and thus the discriminator pulse width. 
Therefore, the CSM algorithm is also a potential method of reducing the effect of charge sharing on time resolution.

The characteristic transition times for a 20~ns nominal exposure shown in Figure~\ref{fig:hs_mode} suggest that sub-100~ns resolution may be possible.
Next, we perform stroboscopic imaging in an optimised setup to estimate the minimum time resolution.

\subsection{Sub-100~ns imaging}
\label{sec5}
To demonstrate the attainable time resolution of the Medipix3 detector, we present an experiment imaging the dynamic illumination source discussed in Section~\ref{sec3}, with the deflection plates driven with a $\pm$10~V 5~MHz square wave.
At this frequency, the illumination source provides a dynamic process with a $\sim$90~ns lifetime.
The experiment was conducted using a similar stroboscopic delay method as that discussed in Section~\ref{sec4}.
The individual frame shutter time was set to 20~ns, with 100k frames contributing to each final image. 
The Medipix3 detector was used in SPM with a threshold voltage of 28~keV to employ the simple architecture of SPM.
The use of CSM would produce similar results but with a longer delay between the electron impact and it being counted.
CSM would also have the benefit of producing images with higher MTFs, which may be important in real experiments, but is not critical for our characterisations of the time response.

\begin{figure}[!ht]
    \centering
    \includegraphics[width=\columnwidth]{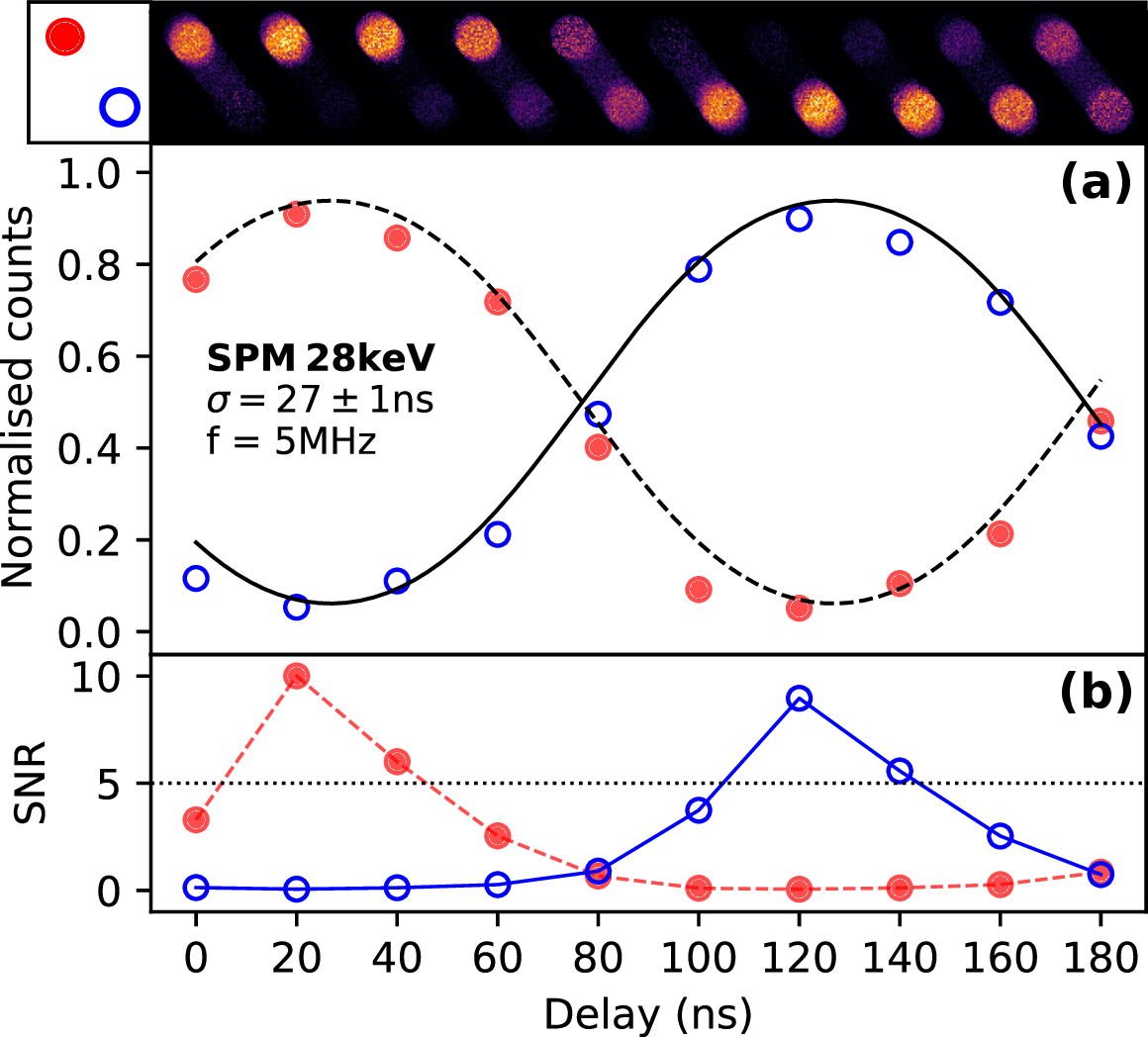}
    \caption{A delay series imaged at 20~ns intervals during a full oscillation of the electron beam at 5~MHz, demonstrating the viability of sub-100~ns imaging. (a) Counts in the two beam positions, and (b) the SNR. The lines in (a) are fits of modelled data to the experimental data (symbols). The summed images are shown in the inset to (a).}  
    \label{fig:SBI}
\end{figure}

The inset to Figure~\ref{fig:SBI}(a) shows a sequence of images taken at delays from 0 to 200~ns in 20~ns increments, encompassing one full cycle of the 5~MHz beam oscillation.
Enlarged versions of the images are shown in Supplemental Fig.~S1.
The edges of the beam appear to broaden slightly along the deflection direction due to the impact of ringing in the driving signal at this time scale, as discussed previously.
The symbols in (a) show the counts in the normalised images in the two beam locations marked by the same symbols adjacent to the inset.
At this frequency of illumination oscillation, the intensity is always at least partially split between spot locations and the intensity profile resembles a modified sinusoidal function.

If one regards the counts in one probe position as signal and all those outwith the disc as noise, then a signal-to-noise ratio (SNR) may be estimated.
The results of this calculation are shown in Figure~\ref{fig:SBI}(b).
The SNR peaks at 10.0 and 9.0 for the top and bottom spot, respectively, significantly exceeding the Rose criterion~\cite{Rose} of an SNR value of 5 (dashed line), confirming that we have stroboscopically imaged a repeated sub-100~ns event.

Further analysis allows us to estimate the minimum time resolution achievable by modelling the intensity profiles of each of the two spots as complementary square waves, each convolved with a Gaussian distribution of standard deviation $\sigma$.
The modelled curves, fitted to the data, are shown as black lines in Figure~\ref{fig:SBI}(a) and give a $\sigma$ of 27$\pm$1~ns.
This time is the quadrature combination of the detector response and that of the illumination source.
However, since the latter is around one order of magnitude smaller than the overall time, correction of the data (-0.14~ns) lies within the error.
Consequently, the time quoted above corresponds to a FWHM defined resolution of ($2\sqrt{2 \ln{(2)}} \sigma$ =) 63$\pm$2~ns.
Alternatively, knowledge of $\sigma$ allows the minimum time resolution, $r$, for imaging an isolated top hat function for a given SNR to be calculated as:
\begin{displaymath}
        r = 2\sqrt{2} \sigma \: \mathrm{erf}^{-1} \left( \frac{\mathrm{SNR}}{\mathrm{SNR}+1} \right).
\end{displaymath}
With an SNR of 5, $r=2.77\sigma$, giving a minimum resolution of 74~ns.
If an SNR of 3 is sufficient, the minimum resolution reduces to 2.30$\sigma = 59$~ns.

\section{Conclusions}
\label{sec6}
We have determined the time resolution of a Medipix3 direct electron detector for 60~keV electrons and outlined the origin of the key limiting factors in the hardware of the device.
Charge sharing between pixels enlarges the point spread function for imaging detectors, reducing the MTF response at high spatial frequencies. 
Our work shows that this charge sharing also impacts the time resolution of the Medipix3 detector.
Increasing the user specified voltage threshold THA in SPM mitigates this effect, improving the time resolution, with values in the region of 59-74~ns demonstrated. 
The use of higher THA in SPM may yield further improvements, but at the expense of substantial reductions in DQE. 
CSM gave comparable results to SPM but it has the advantage of avoiding rejection of electron counts and reduction in MTF.

In TEM imaging, 60~keV is regarded as a low beam energy, often used for the investigation of very thin materials containing light atoms, such as carbon in graphene, with higher beam energies leading to knock-on damage. 
For wider investigation of materials, composed of heavier elements, it is desirable to utilise the higher spatial resolution available at primary beam energies of 200-300~keV.
Of course, electrons at these energies exhibit much larger lateral spreading in the silicon sensor of the Medipix3 with corresponding negative influences on the MTF and temporal resolution. 
Potential performance improvements with these beam energies are being investigated by us through the use of higher atomic number based sensor materials.
A related detector technology, Timepix3 is also being investigated. 
Timepix3 detectors are able to record the time of onset, `time of arrival' and temporal duration of the discriminated pulse. 
Timing information on the former is obtained using a 640~MHz clock and so has a timestamp resolution of 1.56~ns~\cite{Gaspari_JOI_timepix3}.
For imaging with 12~keV X-rays, a practical FWHM time resolution of 19~ns has been demonstrated~\cite{YOUSEF2017639}.
The time resolution of Timepix3 applied in electron microscopy has yet to be determined.
The obtainable practical time resolutions of Timepix3 are likely to be larger than that of more weakly interacting particles, but it should be an improvement on the 59~ns lower limit obtained here for the Medipix3 detector.
The CSM mode of Medipix3, which can improve the DQE and MTF, is absent in Timepix3.
However, using the data driven architecture in Timepix3, it is expected to be possible to track the path of the electron inside the sensor and, as a consequence, determine the primary electron entrance point.
This would potentially allow a similar or greater improvement in the DQE and MTF as a result, but at the expense of orders of magnitude larger data sizes and vastly more complicated analysis.

A further implication from our study relates to STEM imaging. 
Pixelated detectors have become recently employed as ``universal'' detectors~\cite{tate_2016, Hachtel2018,Yang2016, Fang2019} in STEM mode. 
We propose that our results show potential for the application of hybrid pixel detectors to achieve ultimate sensitivity in STEM imaging of beam sensitive materials with the lowest possible electron doses.
The simplest implementation of low-dose STEM is to raster the electron beam as fast as possible across the specimen~\cite{Mittelberger_Ultramicr_2018} with practical pixel dwell times being of the order of $\mathrm{100-200~ns}$. 
We have shown that the Medipix3 detector is capable of registering single electron events with the necessary level of timing accuracy.
However, at the limits of readout speed, Medipix3 is capable of a theoretical maximum frame rate of 24,414 per second (1-bit mode), equivalent to a period of $\mathrm{40.9~{\muup s}}$ per frame, 1 frame per STEM scan pixel.
For an exposure of 10$\times$ the resolution, this corresponds to a duty-cycle of 1.5~\%.
To limit the dose to the specimen while scanning the beam at this lower rate would require the additional coupling of an electrostatic shutter, similar to the deflection plates used in these experiments.
The high temporal resolution that we have measured could then be utilised by triggering from the electrostatic shutter to only acquire electrons in the periods between the unblanking and blanking process.
Alternatively, very recent versions of the Merlin readout system enable temporal gating of the exposure without readout.
This has the potential to allow greater duty-cycles to be achieved by performing hardware-summing, rather than reading out individual frames and then summing them in software.
Allowing sufficient time for the analogue pulse to decay, repeat frequencies over 500~kHz should be achievable.
For the same exposure of 10$\times$ the resolution, this corresponds to a 32~\% duty-cycle.

\section*{Acknowledgements}
This work was supported by EPSRC via the project ``Fast Pixel Detectors: a paradigm shift in STEM imaging'' (Grant reference EP/M009963/1).
G.W.P. received additional support from the EPSRC under grant No. EP/M024423/1.
R.L.'s PhD studentship was supported by EPSRC through the University of Glasgow's doctoral training account.
G.W.P. thanks Kirsty Paton for helpful discussions on the CSM architecture.
All data was processed with Python and the FPD~\cite{fpd_library} and HyperSpy~\cite{hyperspy_library} libraries.
Data files are available at DOI TBA.

% \section*{References}
\bibliographystyle{elsarticle-num}

% \bibliography{refs.bib}

\begin{thebibliography}{10}
\expandafter\ifx\csname url\endcsname\relax
  \def\url#1{\texttt{#1}}\fi
\expandafter\ifx\csname urlprefix\endcsname\relax\def\urlprefix{URL }\fi
\expandafter\ifx\csname href\endcsname\relax
  \def\href#1#2{#2} \def\path#1{#1}\fi

\bibitem{Haider_nature_1998}
{H. Maximilian}, {U. Stephan}, {S. Eugen}, {R. Harald}, {K. Bernd}, {U. Knut},
  {Electron microscopy image enhanced}, Nature 392 (1998) 768.
\newblock \href {http://dx.doi.org/10.1038/33823} {\path{doi:10.1038/33823}}.

\bibitem{pennycook_2006}
S.~Pennycook, M.~Varela, C.~Hetherington, A.~Kirkland, Materials advances
  through aberration-corrected electron microscopy, MRS Bulletin 31~(1) (2006)
  36–43.
\newblock \href {http://dx.doi.org/10.1557/mrs2006.4}
  {\path{doi:10.1557/mrs2006.4}}.

\bibitem{Beacham_JOI_2011}
R.~Beacham, A.~M. Raighne, D.~Maneuski, V.~O'Shea, S.~McVitie, D.~McGrouther,
  {Medipix2/Timepix} detector for time resolved transmission electron
  microscopy, J. Instrum. 6~(12) (2011) C12052.
\newblock \href {http://dx.doi.org/10.1088/1748-0221/6/12/C12052}
  {\path{doi:10.1088/1748-0221/6/12/C12052}}.

\bibitem{PIAZZA_chem_phys_2013}
L.~Piazza, D.~Masiel, T.~LaGrange, B.~Reed, B.~Barwick, F.~Carbone, Design and
  implementation of a fs-resolved transmission electron microscope based on
  thermionic gun technology, J. Chem. Phys. 423 (2013) 79 -- 84.
\newblock \href {http://dx.doi.org/10.1016/j.chemphys.2013.06.026}
  {\path{doi:10.1016/j.chemphys.2013.06.026}}.

\bibitem{Kieft_struc_dyn2_2015}
E.~Kieft, K.~B. Schliep, P.~K. Suri, D.~J. Flannigan, Communication: {Effects}
  of thermionic-gun parameters on operating modes in ultrafast electron
  microscopy, Struct. Dyn. 2~(5) (2015) 051101.
\newblock \href {http://dx.doi.org/10.1063/1.4930174}
  {\path{doi:10.1063/1.4930174}}.

\bibitem{BUCKER20168}
K.~B{\"u}cker, M.~Picher, O.~Cr{\`e}gut, T.~LaGrange, B.~Reed, S.~Park,
  D.~Masiel, F.~Banhart, Electron beam dynamics in an ultrafast transmission
  electron microscope with {Wehnelt} electrode, Ultramicroscopy 171 (2016) 8 --
  18.
\newblock \href {http://dx.doi.org/10.1016/j.ultramic.2016.08.014}
  {\path{doi:10.1016/j.ultramic.2016.08.014}}.

\bibitem{Kuwahara_APL_2016}
M.~Kuwahara, Y.~Nambo, K.~Aoki, K.~Sameshima, X.~Jin, T.~Ujihara, H.~Asano,
  K.~Saitoh, Y.~Takeda, N.~Tanaka, The {Boersch} effect in a picosecond pulsed
  electron beam emitted from a semiconductor photocathode, Appl. Phys. Lett.
  109~(1) (2016) 013108.
\newblock \href {http://dx.doi.org/10.1063/1.4955457}
  {\path{doi:10.1063/1.4955457}}.

\bibitem{FEIST_2017}
A.~Feist, N.~Bach, N.~R. da~Silva, T.~Danz, M.~M{\"o}ller, K.~E. Priebe,
  T.~Domr{\"o}se, J.~G. Gatzmann, S.~Rost, J.~Schauss, S.~Strauch, R.~Bormann,
  M.~Sivis, S.~Sch{\"a}fer, C.~Ropers, Ultrafast transmission electron
  microscopy using a laser-driven field emitter: Femtosecond resolution with a
  high coherence electron beam, Ultramicroscopy 176 (2017) 63 -- 73.
\newblock \href {http://dx.doi.org/10.1016/j.ultramic.2016.12.005}
  {\path{doi:10.1016/j.ultramic.2016.12.005}}.

\bibitem{Feist_nat_2015_nearfield}
A.~Feist, K.~E. Echternkamp, J.~Schauss, S.~V. Yalunin, S.~Sch{\"a}fer,
  C.~Ropers, {Quantum coherent optical phase modulation in an ultrafast
  transmission electron microscope}, Nature 521 (2015) 200.
\newblock \href {http://dx.doi.org/10.1038/nature14463}
  {\path{doi:10.1038/nature14463}}.

\bibitem{Gaolong_sci_rep_2015}
G.~Cao, S.~Sun, Z.~Li, H.~Tian, H.~Yang, J.~Li, {Clocking the anisotropic
  lattice dynamics of multi-walled carbon nanotubes by four-dimensional
  ultrafast transmission electron microscopy}, Sci. Rep. 5 (2015) 8404.
\newblock \href {http://dx.doi.org/10.1038/srep08404}
  {\path{doi:10.1038/srep08404}}.

\bibitem{da_Sliva_2018}
N.~R. da~Silva, M.~M{\"o}ller, A.~Feist, H.~Ulrichs, C.~Ropers, S.~Sch{\"a}fer,
  Nanoscale mapping of ultrafast magnetization dynamics with femtosecond
  lorentz microscopy, Phys. Rev. X 8 (2018) 031052--1 -- 031052--5.
\newblock \href {http://dx.doi.org/10.1103/PhysRevX.8.031052}
  {\path{doi:10.1103/PhysRevX.8.031052}}.

\bibitem{Hassan_nature_2017_electron_dynamics}
M.~T. Hassan, J.~S. Baskin, B.~Liao, A.~H. Zewail, {High-temporal-resolution
  electron microscopy for imaging ultrafast electron dynamics}, Nat. Photon. 11
  (2017) 425.
\newblock \href {http://dx.doi.org/10.1038/nphoton.2017.79}
  {\path{doi:10.1038/nphoton.2017.79}}.

\bibitem{BOSTANJOGLO2000141}
O.~Bostanjoglo, R.~Elschner, Z.~Mao, T.~Nink, M.~Weing{\"a}rtner, Nanosecond
  electron microscopes, Ultramicroscopy 81~(3) (2000) 141 -- 147.
\newblock \href {http://dx.doi.org/10.1016/S0304-3991(99)00180-1}
  {\path{doi:10.1016/S0304-3991(99)00180-1}}.

\bibitem{LAGRANGE_2008_dtem}
T.~LaGrange, G.~H. Campbell, B.~Reed, M.~Taheri, J.~B. Pesavento, J.~S. Kim,
  N.~D. Browning, Nanosecond time-resolved investigations using the in situ of
  dynamic transmission electron microscope {(DTEM)}, Ultramicroscopy 108~(11)
  (2008) 1441 -- 1449.
\newblock \href {http://dx.doi.org/10.1016/j.ultramic.2008.03.013}
  {\path{doi:10.1016/j.ultramic.2008.03.013}}.

\bibitem{BROWNING201223}
N.~Browning, M.~Bonds, G.~Campbell, J.~Evans, T.~LaGrange, K.~Jungjohann,
  D.~Masiel, J.~McKeown, S.~Mehraeen, B.~Reed, M.~Santala, Recent developments
  in dynamic transmission electron microscopy, Curr. Opin. Solid St. M. 16~(1)
  (2012) 23 -- 30.
\newblock \href {http://dx.doi.org/10.1016/j.cossms.2011.07.001}
  {\path{doi:10.1016/j.cossms.2011.07.001}}.

\bibitem{Ji_cathode_2017}
S.~Ji, L.~Piazza, G.~Cao, S.~T. Park, B.~W. Reed, D.~J. Masiel,
  J.~Weissenrieder, Influence of cathode geometry on electron dynamics in an
  ultrafast electron microscope, Struct. Dyn. 4~(5) (2017) 054303--1 --
  054303--18.
\newblock \href {http://dx.doi.org/10.1063/1.4994004}
  {\path{doi:10.1063/1.4994004}}.

\bibitem{Houdellier_Ultramicr_2018}
F.~Houdellier, G.~Caruso, S.~Weber, M.~Kociak, A.~Arbouet, Development of a
  high brightness ultrafast transmission electron microscope based on a
  laser-driven cold field emission source., Ultramicroscopy 186 (2018) 128 --
  138.
\newblock \href {http://dx.doi.org/10.1016/j.ultramic.2017.12.015}
  {\path{doi:10.1016/j.ultramic.2017.12.015}}.

\bibitem{CSM}
R.~Ballabriga, M.~Campbell, E.~H.~M. Heijne, X.~Llopart, L.~Tlustos, The
  {M}edipix3 prototype, a pixel readout chip working in single photon counting
  mode with improved spectrometric performance, IEEE Trans. Nucl. Sci. 54~(5)
  (2007) 1824--1829.
\newblock \href {http://dx.doi.org/10.1109/TNS.2007.906163}
  {\path{doi:10.1109/TNS.2007.906163}}.

\bibitem{Ballabriga_JInstr_2011}
R.~Ballabriga, G.~Blaj, M.~Campbell, M.~Fiederle, D.~Greiffenberg, E.~Heijne,
  X.~Llopart, R.~Plackett, S.~Procz, L.~Tlustos, D.~Turecek, W.~Wong,
  Characterization of the {Medipix3} pixel readout chip, ‎J. Instrum. 6
  (2011) C01052 -- 1--9.
\newblock \href {http://dx.doi.org/10.1088/1748-0221/6/01/C01052}
  {\path{doi:10.1088/1748-0221/6/01/C01052}}.

\bibitem{MIR_UM_2017_medipix}
J.~Mir, R.~Clough, R.~MacInnes, C.~Gough, R.~Plackett, I.~Shipsey, H.~Sawada,
  I.~MacLaren, R.~Ballabriga, D.~Maneuski, V.~O'Shea, D.~McGrouther,
  A.~Kirkland, Characterisation of the {Medipix3} detector for 60 and {80keV}
  electrons, Ultramicroscopy 182 (2017) 44 -- 53.
\newblock \href {http://dx.doi.org/10.1016/j.ultramic.2017.06.010}
  {\path{doi:10.1016/j.ultramic.2017.06.010}}.

\bibitem{PIXDPC}
M.~Krajnak, D.~McGrouther, D.~Maneuski, V.~O. Shea, S.~McVitie, Pixelated
  detectors and improved efficiency for magnetic imaging in {STEM} differential
  phase contrast, Ultramicroscopy 165 (2016) Pages 42--50.
\newblock \href {http://dx.doi.org/10.1016/j.ultramic.2016.03.006}
  {\path{doi:10.1016/j.ultramic.2016.03.006}}.

\bibitem{Rose}
A.~Rose, A unified approach to the performance of photographic film, television
  pickup tubes and the human eye, Journal of the SMPTE 47~(4) (1946) 273.
\newblock \href {http://dx.doi.org/10.5594/J12772} {\path{doi:10.5594/J12772}}.

\bibitem{Llopart_IEEE_medipix2_2002}
X.~Llopart, M.~Campbell, R.~Dinapoli, D.~S. Segundo, E.~Pernigotti, Medipix2:
  {A} 64-k pixel readout chip with 55$\muup$m square elements working in single
  photon counting mode, IEEE Trans. Nucl. Sci. 49~(5) (2002) 2279--2283.
\newblock \href {http://dx.doi.org/10.1109/TNS.2002.803788}
  {\path{doi:10.1109/TNS.2002.803788}}.

\bibitem{casino_2007}
D.~Drouin, A.~R. Couture, D.~Joly, X.~Tastet, V.~Aimez, R.~Gauvin, {CASINO
  V2.42—A} fast and easy-to-use modeling tool for scanning electron
  microscopy and microanalysis users, Scanning 29~(3) (2007) 92--101.
\newblock \href {http://dx.doi.org/10.1002/sca.20000}
  {\path{doi:10.1002/sca.20000}}.

\bibitem{Horswell_2011_joi_100MHz}
I.~Horswell, E.~N. Gimenez, J.~Marchal, N.~Tartoni, A {Medipix3} readout system
  based on the {N}ational {I}nstruments {FlexRIO} card and using the {LabVIEW}
  programming environment, J. Instrum. 6~(01) (2011) C01028.
\newblock \href {http://dx.doi.org/10.1088/1748-0221/6/01/C01028}
  {\path{doi:10.1088/1748-0221/6/01/C01028}}.

\bibitem{Plackett_2013_joi_jitter}
R.~Plackett, I.~Horswell, E.~N. Gimenez, J.~Marchal, D.~Omar, N.~Tartoni,
  Merlin: a fast versatile readout system for {Medipix3}, ‎J. Instrum. 8~(01)
  (2013) C01038.
\newblock \href {http://dx.doi.org/10.1088/1748-0221/8/01/C01038}
  {\path{doi:10.1088/1748-0221/8/01/C01038}}.

\bibitem{Hayashida2007505}
M.~Hayashida, T.~Nomaguchi, Y.~Kimura, Y.~Takai, Development of
  computer-assisted minimal-dose system with beam blanker for {TEM}, Micron
  38~(5) (2007) 505 -- 512.
\newblock \href {http://dx.doi.org/10.1016/j.micron.2006.07.024}
  {\path{doi:10.1016/j.micron.2006.07.024}}.

\bibitem{Ohtsuki1975163}
M.~Ohtsuki, E.~Zeitler, Minimal beam exposure with a field emission source,
  Ultramicroscopy 1~(2) (1975) 163--165.
\newblock \href {http://dx.doi.org/10.1016/S0304-3991(75)80021-0}
  {\path{doi:10.1016/S0304-3991(75)80021-0}}.

\bibitem{McVitie_Ultramicr}
S.~McVitie, D.~McGrouther, S.~McFadzean, D.~MacLaren, K.~O’Shea, M.~Benitez,
  Aberration corrected {Lorentz} scanning transmission electron microscopy,
  Ultramicroscopy 152 (2015) 57 -- 62.
\newblock \href {http://dx.doi.org/10.1016/j.ultramic.2015.01.003}
  {\path{doi:10.1016/j.ultramic.2015.01.003}}.

\bibitem{Richardson:72}
W.~H. Richardson, Bayesian-based iterative method of image restoration, J. Opt.
  Soc. Am. 62~(1) (1972) 55--59.
\newblock \href {http://dx.doi.org/10.1364/JOSA.62.000055}
  {\path{doi:10.1364/JOSA.62.000055}}.

\bibitem{Lucy:1974yx}
L.~B. Lucy, {An iterative technique for the rectification of observed
  distributions}, Astron. J. 79 (1974) 745--754.
\newblock \href {http://dx.doi.org/10.1086/111605} {\path{doi:10.1086/111605}}.

\bibitem{Gaspari_JOI_timepix3}
M.~D. Gaspari, J.~Alozy, R.~Ballabriga, M.~Campbell, E.~Fr{\"o}jdh,
  J.~Idarraga, S.~Kulis, X.~Llopart, T.~Poikela, P.~Valerio, W.~Wong, Design of
  the analog front-end for the {Timepix3} and {Smallpix} hybrid pixel detectors
  in 130 nm {CMOS} technology, J. Instrum. 9~(01) (2014) C01037.
\newblock \href {http://dx.doi.org/10.1088/1748-0221/9/01/C01037}
  {\path{doi:10.1088/1748-0221/9/01/C01037}}.

\bibitem{YOUSEF2017639}
H.~Yousef, G.~Crevatin, E.~N. Gimenez, I.~Horswell, D.~Omar, N.~Tartoni,
  Timepix3 as {X-ray} detector for time resolved synchrotron experiments,
  Nuclear Instruments and Methods in Physics Research Section {A}:
  Accelerators, Spectrometers, Detectors and Associated Equipment 845 (2017)
  639 -- 643, proceedings of the Vienna Conference on Instrumentation 2016.
\newblock \href {http://dx.doi.org/10.1016/j.nima.2016.04.075}
  {\path{doi:10.1016/j.nima.2016.04.075}}.

\bibitem{tate_2016}
M.~W. Tate, P.~Purohit, D.~Chamberlain, K.~X. Nguyen, R.~Hovden, C.~S. Chang,
  P.~Deb, E.~Turgut, J.~T. Heron, D.~G. Schlom, D.~C. Ralph, G.~D. Fuchs, K.~S.
  Shanks, H.~T. Philipp, D.~A. Muller, S.~M. Gruner, High dynamic range pixel
  array detector for scanning transmission electron microscopy, Microsc.
  Microanal. 22~(1) (2016) 237–249.
\newblock \href {http://dx.doi.org/10.1017/S1431927615015664}
  {\path{doi:10.1017/S1431927615015664}}.

\bibitem{Hachtel2018}
J.~A. Hachtel, J.~C. Idrobo, M.~Chi, Sub-{\aa}ngstrom electric field
  measurements on a universal detector in a scanning transmission electron
  microscope, Adv. Struct. and Chem. Imaging 4~(1) (2018) 10.
\newblock \href {http://dx.doi.org/10.1186/s40679-018-0059-4}
  {\path{doi:10.1186/s40679-018-0059-4}}.

\bibitem{Yang2016}
H.~Yang, R.~N. Rutte, L.~Jones, M.~Simson, R.~Sagawa, H.~Ryll, M.~Huth, T.~J.
  Pennycook, M.~Green, H.~Soltau, Y.~Kondo, B.~G. Davis, P.~D. Nellist,
  {Simultaneous atomic-resolution electron ptychography and {Z-contrast}
  imaging of light and heavy elements in complex nanostructures}, Nat. Commun.
  7 (2016) 12532.
\newblock \href {http://dx.doi.org/10.1038/ncomms12532}
  {\path{doi:10.1038/ncomms12532}}.

\bibitem{Fang2019}
S.~Fang, Y.~Wen, C.~S. Allen, C.~Ophus, G.~G.~D. Han, A.~I. Kirkland,
  E.~Kaxiras, J.~H. Warner, Atomic electrostatic maps of {1D} channels in {2D}
  semiconductors using {4D} scanning transmission electron microscopy, Nat.
  Commun. 10~(1) (2019) 1127.
\newblock \href {http://dx.doi.org/10.1038/s41467-019-08904-9}
  {\path{doi:10.1038/s41467-019-08904-9}}.

\bibitem{Mittelberger_Ultramicr_2018}
A.~Mittelberger, C.~Kramberger, J.~C. Meyer, Software electron counting for
  low-dose scanning transmission electron microscopy, Ultramicroscopy 188
  (2018) 1--7.
\newblock \href {http://dx.doi.org/10.1016/j.ultramic.2018.02.005}
  {\path{doi:10.1016/j.ultramic.2018.02.005}}.

\bibitem{fpd_library}
{FPD: Fast pixelated detector data storage, analysis and visualisation.
  https://gitlab.com/fpdpy/fpd (accessed May 20, 2019)}.

\bibitem{hyperspy_library}
F.~de~la Pe{\~n}a, T.~Ostasevicius, V.~T. Fauske, P.~Burdet, E.~Prestat,
  P.~Jokubauskas, M.~Nord, M.~Sarahan, K.~E. MacArthur, D.~N. Johnstone,
  J.~Taillon, J.~Caron, V.~Migunov, T.~Furnival, A.~Eljarrat, S.~Mazzucco,
  T.~Aarholt, M.~Walls, T.~Slater, F.~Winkler, B.~Martineau, G.~Donval,
  R.~McLeod, E.~R. Hoglund, I.~Alxneit, I.~Hjorth, T.~Henninen, L.~F. Zagonel,
  A.~Garmannslund, 5ht2, hyperspy/hyperspy: {HyperSpy} 1.3.1 (Apr. 2018).
\newblock \href {http://dx.doi.org/10.5281/zenodo.1221347}
  {\path{doi:10.5281/zenodo.1221347}}.

\end{thebibliography}

\end{document}